%% file: article_4.tex
\documentclass[
10pt, 
a4paper, 
oneside, 
headinclude,footinclude, 
BCOR5mm, 
]{scrartcl}

\input{structure.tex} 

\title{\normalfont\spacedallcaps{A Light CNN for detecting COVID-19 from CT scans of the chest*}} 

\author{\spacedlowsmallcaps{Matteo Polsinelli \textsuperscript{1}, Luigi Cinque\textsuperscript{2} \& Giuseppe Placidi\textsuperscript{1}}} 

\date{} 

\begin{document}


\renewcommand{\sectionmark}[1]{\markright{\spacedlowsmallcaps{#1}}} 
\lehead{\mbox{\llap{\small\thepage\kern1em\color{halfgray} \vline}\color{halfgray}\hspace{0.5em}\rightmark\hfil}} 

\pagestyle{scrheadings} 


\maketitle 

\setcounter{tocdepth}{2} 





\section*{Abstract} 

COVID-19 is a world-wide disease that has been declared as a pandemic by the World Health Organization. Computer Tomography (CT) imaging of the chest seems to be a valid diagnosis tool to detect COVID-19 promptly and to control the spread of the disease. Deep Learning has been extensively used in medical imaging and convolutional neural networks (CNNs) have been also used for classification of CT images. We propose a light CNN design based on the model of the SqueezeNet, for the efficient discrimination of COVID-19 CT images with other CT images (community-acquired pneumonia and/or healthy images). On the tested datasets, the proposed modified SqueezeNet CNN achieved 83.00\% of accuracy, 85.00\% of sensitivity, 81.00\% of specificity, 81.73\% of precision and 0.8333 of F1Score in a very efficient way (7.81 seconds medium-end laptot without GPU acceleration). Besides performance, the average classification time is very competitive with respect to more complex CNN designs, thus allowing its usability also on medium power computers. In the next future we aim at improving the performances of the method along two directions: 1) by increasing the training dataset (as soon as other CT images will be available); 2) by introducing an efficient pre-processing strategy.

\let\thefootnote\relax\footnotetext{\textsuperscript{1} \textit{A2VI Lab, Dept. of Life, Health
and Environmental Sciences, University of L’Aquila, Via Vetoio, L’Aquila 67100, Italy}}

\let\thefootnote\relax\footnotetext{\textsuperscript{2} \textit{Dept. Computer Science, Via Salaria, Sapienza University, Rome, Italy}}

\let\thefootnote\relax\footnotetext{* \textit{Manuscript submitted to Pattern Recognition Letters}}


\newpage 


\section{Introduction}

Coronavirus (COVID‐19) is a world-wide disease that has been declared as a pandemic by the World Health Organization on 11th March 2020. To date, more than two million people have been infected and more than 160 thousand died. A quick diagnosis is fundamental to control the spread of the disease and increases the effectiveness of medical treatment and, consequently, the chances of survival without the necessity of intensive and sub-intensive care. This is a crucial point because hospitals have  limited availability of equipment for intensive care. Viral nucleic acid detection using real-time polymerase chain reaction (RT-PCR) is the accepted standard diagnostic method. However, many countries are unable to provide the sufficient RT-PCR due to the fact that the disease is very contagious. So, only people with evident symptoms are tested. Moreover, it takes several hours to furnish a result. Therefore,  faster and reliable screening techniques that could be further confirmed by the PCR test (or replace it) are required.

Computer tomography (CT) imaging seems to be a valid alternative to detect COVID-19 \cite{chua2020role} with a higher sensitivity \cite{fang2020sensitivity} (up to 98\% compared with 71\% of RT-PCR). CT is likely to become increasingly important for the diagnosis and management of COVID-19 pneumonia, considering the continuous increments  in global cases. Early research shows a pathological pathway that might be amenable to early CT detection, particularly if the patient is scanned 2 or more days after developing symptoms\cite{chua2020role}. Nevertheless, the main bottleneck that radiologists experience in analysing radiography images is the visual scanning of small details. Moreover, a large number of CT images have to be evaluated in a very short time thus increasing the probability of misclassifications. This justifies the use of intelligent approaches that can automatically classify CT images of the chest.

\begin{figure}[!t]
\centerline{\includegraphics[width=\columnwidth]{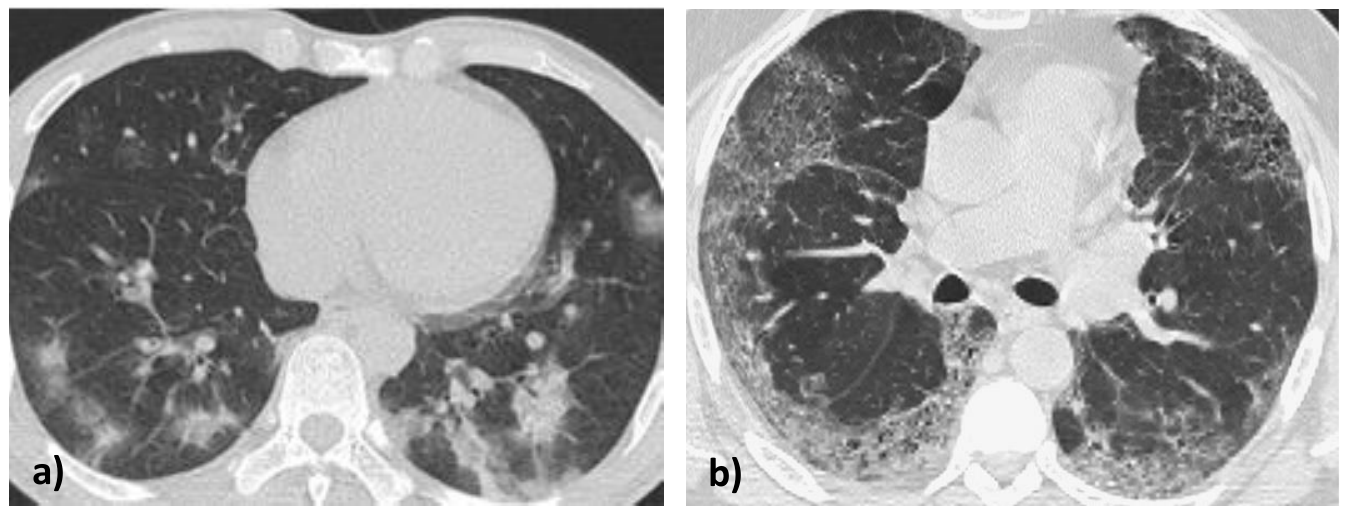}}
\caption{Images extracted from dataset \cite{zhao2020covid}. A COVID-19 image (a) and a not COVID-19 image also containing inflammations (b).}
\label{figcovi}
\end{figure}

Deep Learning methods have been extensively used in medical imaging. In particular, convolutional neural networks (CNNs) have been used both for classification and segmentation problems, also of CT images~\cite{xu2019efficient}. Though CNNs have demonstrated promising performance in this kind of applications, they require a lot of data to be correctly trained. In fact, CT images of the lungs can be easily misclassified, especially when both contain damages due to pneumonia, referred due to different causes (Figure \ref{figcovi}).
Until now, there are limited datasets for COVID-19 and those available contain a limited number of CT images. For this reason, during the training phase it is necessary to avoid/reduce overfitting (that means the CNN is not learning the discriminant features of COVID-19 CT scans but only memorizing it). Another critical point is that CNN inference requires a lot of computational power. In fact, usually CNNs are executed on particularly expensive GPUs equipped with specific hardware acceleration systems. Anyway, expensive GPUs are still the exception rather than the norm in normal computing clusters that usually are CPU based \cite{vanhoucke2011improving}. Even more, this type of machines could not be available be available in hospitals, especially in emergency situations and/or in developing countries.

In the present work, we aim at obtaining acceptable performances for an automatic method in recognizing COVID-19 CT images of lungs while, at the same time, dealing with reduced datasets for training and validation and reducing the computational overhead imposed by more complex automatic systems. 

For this reason, in this work we started from the model of the SqueezeNet CNN, because it is able to reach the same accuracy of modern CNNs but with fewer parameters\cite{iandola2016squeezenet}. Moreover, in a recent benchmark \cite{bianco2018benchmark}, SqueezeNet has achieved the best accuracy density (accuracy divided by number of parameters) and the best inference time.

To date, some works on COVID-19 detection by CT images are being published \cite{wang2020deep, li2020artificial, xu2020deep}. All these works use heavy CNNs (respectively resnet, inception and resnet) adapted to improve accuracy.

In this work we developed, trained and tested a light CNN (based on the SqueezeNet) to discriminate between COVID-19 and community-acquired pneumonia and/or healthy CT images. The hyper-parameters have been optimized with Bayesian method on two datasets\cite{zhao2020covid, italianDataset}. In addition, class activation mapping (CAM)\cite{zhou2016learning} has been used to understand which parts of the image are relevant for the CNN to classify it and to check that no over-fitting occurs.

The paper is structured as follow: in the next section (Materials and Methods) the datasets organization, the used processing equipment and the proposed methodology are presented; section 3 contains Results and Discussion, including a comparison with recent works on the same argument; finally section 4 concludes the paper and proposes future improvements.


\section{Methods}

\subsection{Datasets organization}
The datasets used therein are the Zhao et al. dataset\cite{zhao2020covid} and the Italian dataset\cite{italianDataset}. The Zhao et al. dataset\cite{zhao2020covid} is composed by 360 CT scans of COVID-19 subjects and 397 CT scans of other kinds of illnesses and/or healthy subjects. The italian dataset is composed of 100 CT scans of COVID-19. These datasets are continuously updating and their images is raising at the same time. 
In this work we used two different arrangements of the datasets, one in which data from both datasets are used separately and the other containing data mixed by both datasets. The first arrangement contains two different test datasets (Test-1 and Test-2). In fact, the Zhao dataset is used alone and divided in train, validation and Test-1. The italian dataset is integrated into a second test dataset, Test-2 (Table \ref{tab:my-table1}), while the  Zhao dataset is always used in train, validation and Test-2 (in Test-2, the not COVID-19 images of the Zhao dataset are the same of Test-1). The first arrangement is used to check if, even with a small training dataset, it is possible to train a CNN capable to work well also on a completely different and new dataset (the italian one).
In the second arrangement, both datasets are mixed as indicated in Table \ref{tab:my-table2}. In this arrangement the number of images from the italian dataset used to train, validate and Test-1 are 60, 20 and 20, respectively. The second arrangement represents a more realistic case in which both datasets are mixed to increase as possible the training dataset (at the expenses of a Test-2 which, in this case, is absent). 
In both arrangements, the training dataset has been augmented with the following transformations: a rotation (with a random angle between 0 and 90 degrees), a scale (with a random value between 1.1 and 1.3) and addition of gaussian noise to the original image.


\begin{table*}[!t]
\caption{Dataset arrangement 1}
\centering
\begin{tabular}{|p{2cm}|p{2cm}|p{2.5cm}|p{2cm}|p{1cm}|}
\cline{2-5}
\multicolumn{1}{l|}{}                     & \textbf{COVID-19} & \textbf{Not COVID-19} & \textbf{Data Augm.}                                                                         & \textbf{Total} \\ \hline
\multicolumn{1}{|c|}{\textbf{Train}}      & 191               & 191                   & \multicolumn{1}{l|}{x4} & 1528           \\ \hline
\multicolumn{1}{|c|}{\textbf{Validation}} & 60                & 58                    & No                                                                                                  & 118            \\ \hline
\multicolumn{1}{|c|}{\textbf{Test-1}}     & 98                & 95                    & No                                                                                                  & 193            \\ \hline
\multicolumn{1}{|c|}{\textbf{Test-2}}     & 100               & 95                    & No                                                                                                  & 195            \\ \hline
\end{tabular}

\label{tab:my-table1}
\end{table*}

\begin{table*}[!t]
\caption{Dataset arrangement 2}
\centering
\begin{tabular}{|p{2cm}|p{2cm}|p{2.5cm}|p{2cm}|p{1cm}|}
\cline{2-5}
\multicolumn{1}{l|}{} &
  \textbf{COVID-19} &
  \textbf{Not COVID-19} &
  \textbf{Data Augm.} &
  \textbf{Total} \\ \hline
\multicolumn{1}{|c|}{\textbf{Train}} &
  251 &
  191 &
  \multicolumn{1}{l|}{x4} &
  1768 \\ \hline
\multicolumn{1}{|c|}{\textbf{Validation}} & 80  & 58 & No & 138 \\ \hline
\multicolumn{1}{|c|}{\textbf{Test-1}}     & 108 & 95 & No & 203 \\ \hline
\end{tabular}
\label{tab:my-table2}
\end{table*}
 
\subsection{Computational resources}

For the numerical of the proposed CNNs we used two hardware systems: 
1) a high level computer with CPU Intel Core i7-67100, RAM 32 GB and GPU Nvidia GeForce GTX 1080 8 GB dedicated memory; 2) a low level laptot with CPU Intel Core i5 processor, RAM 8 GB and no dedicated GPU. The first is used for hyperparameters optimization and to train, validate and test the CNNs; the second is used just for test in order to demonstrate the computational efficiency of the proposed solution. 

In both cases we used the development environment Matlab 2020a. Matlab integrates powerful toolboxes for the design of neural networks. Moreover, with Matlab it is possible to export the CNNs in an open source format called “ONNX”, useful to share the CNNs with research community. When used the high level computer is used, the GPU acceleration is enabled in Matlab environment, based on the technology Nvida Cuda Core provided by the GPU that allows parallel computing. In this way we speed up the prototyping of the CNNs. When final tests are performed on the low level hardware, no GPU acceleration is used.
\begin{figure}[!t]
\centerline{\includegraphics[width=0.60\columnwidth]{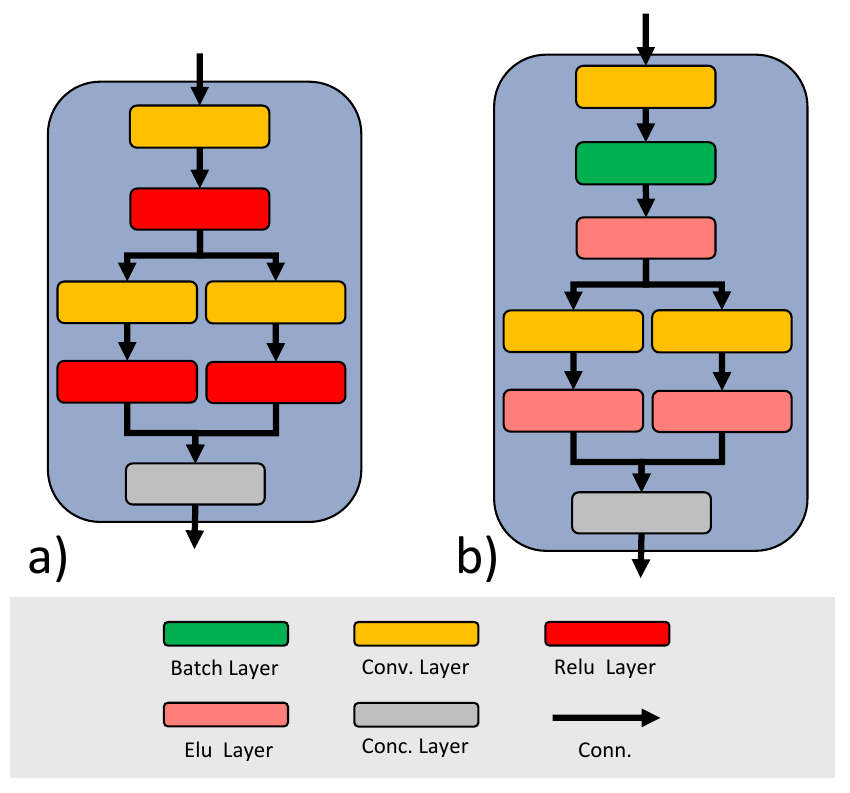}}
\caption{The classic Fire Module of the SqueezeNet (a). Proposed modification to the Fire Module of the SqueezeNet performed to improve convergence and to reduce overfitting (b).}
\label{fig1}
\end{figure}

\begin{figure*}[!t]
\centerline{\includegraphics[width=\columnwidth]{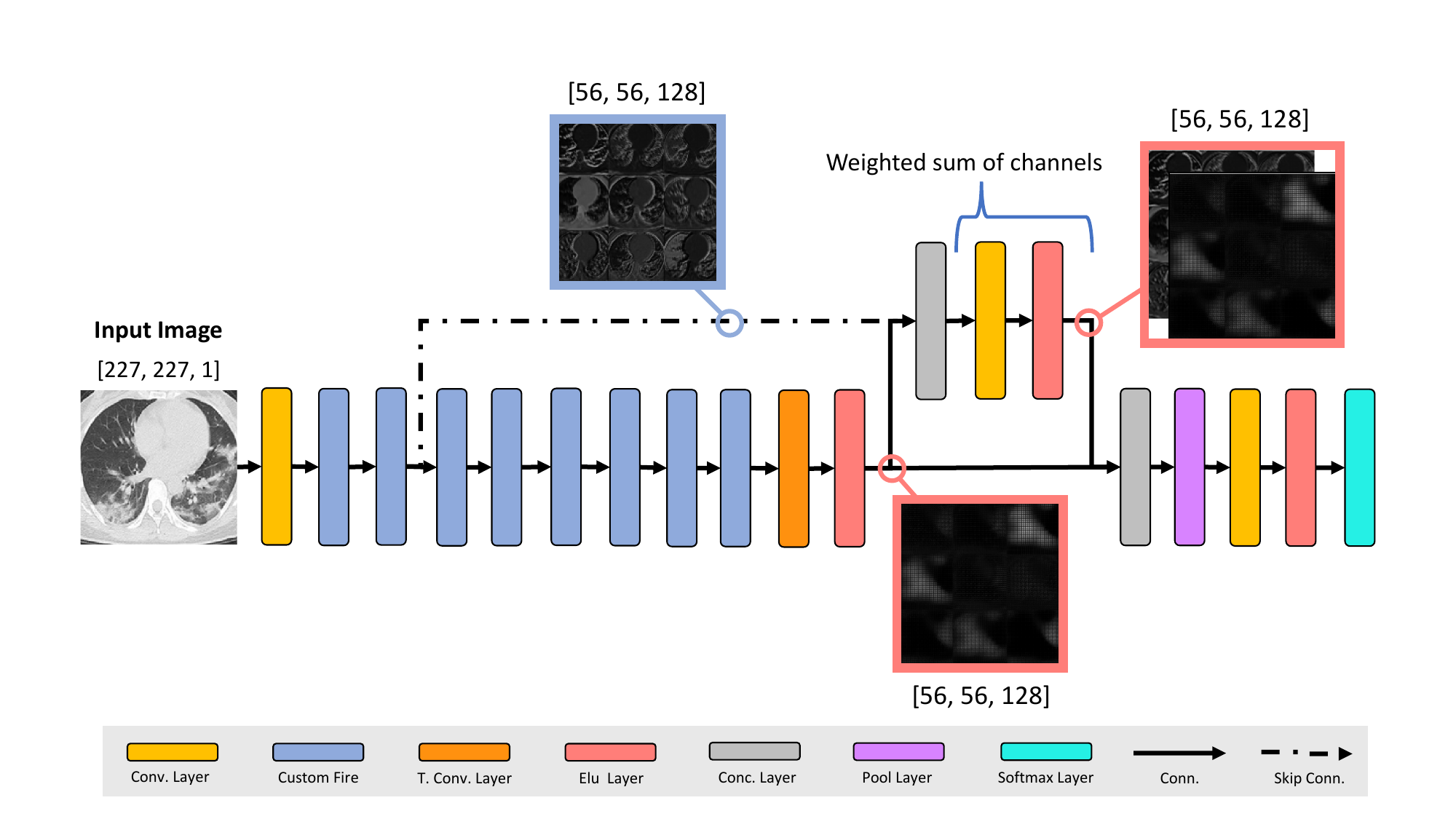}}
\caption{The proposed custom CNN. Spatial information contained in the feature maps from the second Custom Fire Module are weighted with the feature maps of the last Custom Fire Module.}
\label{figsqueeze}
\end{figure*}

\subsection{CNN design}

The SqueezeNet is capable of achieving the same level of accuracy of others, more complex, CNNs designs which have a huge number of layers and parameters\cite{iandola2016squeezenet}. For example, SqueezeNet can achieve the same accuracy of Alex-Net \cite{krizhevsky2012imagenet} on the ImageNet dataset\cite{deng2009imagenet} with 50X fewer parameters and a model size of less than 0.5MB\cite{iandola2016squeezenet}.
The SqueezeNet is composed of \textit{blocks} called "Fire Module". As shown in Figure \ref{fig1}.a,  each block is composed of a \textit{squeeze} convolution layer (which has 1x1 filters) feeding an \textit{expanding section} of two convolution layers with 1x1 and 3x3 filters, respectively. Each convolution layer 
is followed by a ReLu layer. The ReLu layers output of the expanding section are concatenated with a Concatenation layer. To improve the training convergence and to reduce overfitting we added a Batch Normalization layer between the squeeze convolution layer and the ReLu layer (Figure \ref{fig1}.b). Each Batch Normalization layer adds 30\% of computation overhead and for this reason we chose to add them only before the expanding section in order to make it more effective while, at the same time, limiting their number. Moreover, we replaced all the Relu layers with Elu layers because,  from literature \cite{clevert2015fast}, ELUs networks without Batch Normalization significantly outperform ReLU networks with Batch Normalization.

The SqueezeNet has 8 Fire Modules in cascade configuration. Anyway, two more complex architectures exist: one with simple and another with complex bypass. The simple bypass configuration consists in 4 skip connections added between Fire Module 2 and Fire Module 3, Fire Module 4 and Fire Module 5, Fire Module 6 and Fire Module 7 and, finally, between Fire Module 8 and Fire Module 9. The complex bypass added 4 more skip connections (between the same Fire Modules) with a convolutional layer of filter size 1x1. From the original paper \cite{iandola2016squeezenet} it seems that the better accuracy is achieved by the simpler bypass configuration. For this reason, in this work we test both  SqueezeNet without any bypass (to have the most efficient model) and with simple bypass (to have the most accurate model), while complex bypass configuration is not considered. 

Besides, we propose also a further modify CNN (Figure \ref{figsqueeze}) based on the SqueezeNet without any bypass. Moreover, we added a Transpose Convolutional Layer to the last Custom Fire Module that expands the feature maps 4 times along width and height dimensions. These feature maps are concatenated in depth with the feature maps from the second Custom Fire Module through a skip connection. Weighted sum is performed between them with a Convolution Layer with 128 filters of size 1x1. Finally all the feature map are concatenated in depth and averaged with a Global Average Pool Layer. This design allows to combine spatial information (early layers) and features information ( last layers) to improve the accuracy.

\subsection{Hyperparameters tuning}
Since we are using a light CNN to classify, the optimization of the training phase is crucial to achieve good results with a limited number of parameters.
The training phase of a CNN is highly correlated with settings hyperparameters. Hyperparameters are different from model weights. The former are calculated before the training phase, whereas the latter are optimised during the training phase. Setting of hyperparameters is not trivial and different strategies can be adopted. A first way is to select hyperparameters manually though it would be preferable to avoid it because the number of different configurations is huge. For the same reason, approaches like grid search do not use do not use past evaluations: a lot of time has to be spent for evaluating bad hyperparameters configurations. Instead, Bayesian approaches, by using past evaluation results to build a \textit{surrogate} probabilistic model mapping hyperparameters to a probability of a score on the objective function, seem to work better.

In this work we used Bayesian optimization for the following hyper-parameters:
\begin{enumerate}
\item \textbf{Initial Learning Rate}: the rate used for updating weights during the training time;
\item \textbf{Momentum}: this parameter influences the weights update taking into consideration the update value of the previous iteration;
\item \textbf{L2-Regularization}: a regularization term for the weights to the loss function in order to reduce over-fitting. 
\end{enumerate}

\section{Results and Discussion}

\subsection{Experiments organization and hyperparameters optimization}
For each dataset arrangement we organized 4 experiments in which we tested different CNN models, transfer learning and the effectiveness of data augmentation. For each experiment, 30 different attempts (with Bayesian method) have been made with different set of hyper-parameters (Initial Learning Rate, Momentum, L2-Regularization). For each attempt, the CNN model has been trained for 20 epochs and evaluated by the accuracy results calculated on the validation dataset.
The experiments, all performed on the augumented dataset were:

\begin{enumerate}
\item SqueezeNet without bypass and transfer learning;
\item SqueezeNet with simple bypass but without transfer learning;
\item SqueezeNet with simple bypass and transfer learning;
\item the proposed CNN.
\end{enumerate}

\begin{table}[]
\caption{Results on the dataset arrangement 1}
\centering
\begin{tabular}{|p{1cm}|p{1.8cm}|p{1.8cm}|p{2cm}|p{1.8cm}|p{1.8cm}|}
\hline
\textbf{Exp.} & \textbf{Obs. Acc.} & \textbf{Est. Acc.} & \textbf{Learn. Rate} & \textbf{Mom.} & \textbf{L2-Reg.}     \\ \hline 
1 & 88.30\%  & 82.26\% & 0.074516  & 0.58486  & 1.6387e-07 \\ \hline
2 & 85.76\%  & 82.42\% & 0.011358   & 0.97926 & 3.684e-08  \\ \hline
3 & 85.76\%  & 80.58\% & 0.00070093 & 0.96348 & 1.0172e-12 \\ \hline
\textbf{4} & \textbf{89.85}\%  & \textbf{87.27}\% & \textbf{0.007132} & \textbf{0.87589} & \textbf{0.9532e-06} \\ \hline
\end{tabular}
\label{tab:my-table3}
\end{table}

\begin{table*}[!t]
\centering
\caption{Results on the dataset arrangement 2}
\begin{tabular}{|p{1cm}|p{1.8cm}|p{1.8cm}|p{2cm}|p{1.8cm}|p{1.8cm}|}
\hline
\textbf{Exp.} & \textbf{Obs. Acc.} & \textbf{Est. Acc.} & \textbf{Learn. Rate} & \textbf{Mom.} & \textbf{L2-Reg.}     \\ \hline
1 & 86,84\% & 82.11\% & 0.00010091 & 0.70963 & 2.2153e-11 \\ \hline
2 & 85.36\%  & 81.53\% & 0.086175    & 0.59589  & 7.5468e-09  \\ \hline
3 & 84.44\%  & 80.22\% & 0.0016053 & 0.86453 & 1.0048e-10 \\ \hline
\textbf{4} & \textbf{87.56}\%  & \textbf{85.87}\% & \textbf{0.089642} & \textbf{0.84559} & \textbf{0.5895e-07} \\ \hline
\end{tabular}
\label{tab:my-table4}
\end{table*}

Regarding the arrangement 1, the results of the experiments are reported in Table \ref{tab:my-table3}. 
For a better visualization of the results, we report just the the best accuracy calculated with respect to all the attempts, the accuracy estimated by the objective function at the end of all attempts and the values of the hyperparameters.
The best accuracy value is achieved with the experiment \#4. Both observed and estimated accuracy are the highest between all the experiments. Regarding the original paper of the SqueezeNet \cite{iandola2016squeezenet}, it seems that there is not a relevant difference between the model without bypass and with bypass. It is also interesting to note that use transfer learning (experiment \#3) from the original weights of the SqueezeNet does not have a relevant effect. Regarding the dataset arrangement 2, the results of the experiments are shown in Table \ref{tab:my-table4}. The experiment \#4 is still the best one, though experiment \#1 is closer in terms of observed accuracy. However, we did not expect such a difference between the learning rate of experiment \#4 of Table \ref{tab:my-table3} and Table \ref{tab:my-table4}. Moreover, also the L2-Regularization changed a lot. It suggests that the CNN  trained/validated on the dataset arrangement 1 (that we call CNN-1) has a different behavior with respect to the CNN trained/validated on dataset arrangement 2 (that we call CNN-2).

However, the results shown in Table \ref{tab:my-table3} and Table \ref{tab:my-table4} suggest that the proposed CNN achieves better results when compared to different configurations of the original SqueezeNet.

\subsection{Training, Validation and Test}

Both CNN-1 and CNN-2 have been trained for more 20 epochs, with a Learning Rate drop of 0.8 every 5 epochs. 
After that, both CNNs have been evaluated with the respective Test-1 dataset with the following benchmark metrics: Accuracy (measures the correct predictions of the CNN), Sensitivity (measures the positives that are correctly identified),
Specificity (measures the negatives that are correctly identified), Precision (measures the proportion of positive identification that is actually correct) and
F1Score(measures the balance between Precision and Recall).


The results, shown in Table \ref{tab:my-table5}, confirm the hypothesis of the previous section: CNN-1 and CNN-2 have a different behavior. This is clearly understandable by taking into account the Sensitivity and Specificity values. The CNN-1 has higher Specificity (0.85 against 0.81) and that means that is capable to better recognize not COVID-19 images. The CNN-2 has higher Sensitivity (0.8500 against 0.7900) and that means that is capable to better recognize COVID-19 images. 

Regarding the application of CNN-1 on Test-2, the results are frustrating. The accuracy reaches just 0.5024 because the CNN is capable only to recognize well not COVID-19 images (precision is 0.80) but has very poor performance on COVID-19 images (sensitivity = 0.1900). As affirmed before, the analyses of Test-2 is very hard if we do not use a larger dataset of images.

\begin{table}[]
\center
\caption{CNN-1 and CNN-2 performances}
\begin{tabular}{|c|c|c|c|c|c|}
\hline
\textbf{CNN} & \textbf{Acc.} & \textbf{Sens.} & \textbf{Spec.} & \textbf{Prec.} & \textbf{F1Score} \\ \hline
CNN-1        &  0.8200           & 0.7900               &  0.8500 & 0.8404             &  0.8144          \\ \hline
CNN-2        & 0.8300            & 0.8500               & 0.8100 &   0.8173 & 0.8333         \\ \hline
\end{tabular}
\label{tab:my-table5}
\end{table}

In order to deeply understand the behaviour of CNN-1 and CNN-2 we used CAM \cite{zhou2016learning}, that gives a visual explanations of the predictions of convolutional neural networks. This is useful to figure out what each CNN has learned and which part of the input of the network is responsible for the classification. It can be useful to identify biases in the training set and to increase model accuracy. With CAM it is also possible to understand if the CNNs are overfitting. In fact, if the network has high accuracy on the training set, but low accuracy on the Test set, CAM helps to verify if the CNN is basing its predictions on the relevant features of the images or on the background. To this aim, we expect that the activations maps are focused on the lungs and especially on those parts affected by COVID-19 (lighter regions with respect to healthy, darker, zones of the lungs). 

\begin{table*}[!t]
\caption{Comparison with previous works}
\centering
\begin{tabular}{|p{1.7cm}|p{1.3cm}|p{1.5cm}|p{1.5cm}|p{1.5cm}|p{1.5cm}|p{1cm}|}
\hline
\textbf{Works} &
  \textbf{\begin{tabular}[c]{@{}c@{}}Image \\ Preproc\end{tabular}} &
  \textbf{\begin{tabular}[c]{@{}c@{}}Accuracy\\ (\%)\end{tabular}} &
  \textbf{\begin{tabular}[c]{@{}c@{}}Sensitivity\\ (\%)\end{tabular}} &
  \textbf{\begin{tabular}[c]{@{}c@{}}Specificity\\ (\%)\end{tabular}} &
  \textbf{\begin{tabular}[c]{@{}c@{}}Precision\\ (\%)\end{tabular}} &
  \textbf{\begin{tabular}[c]{@{}c@{}}F1\\ Score\end{tabular}} \\ \hline    Wang et al.\cite{wang2020deep} & No  & 73.1  & 67    & 76    & 61    & 0.63   \\ \hline
Xu et al. \cite{xu2020deep}  & Yes & -     & 86.7  & -     & 81.3  & 0.839  \\ \hline
Li et al. \cite{li2020artificial}         & Yes & -     & 90    & 96    & -     & -      \\ \hline
The proposed CNN           & No  & 83.00 & 85.00 & 81.00 & 81.73 & 83.33 \\ \hline
\end{tabular}
\label{tab:my-table6}
\end{table*}

\begin{figure}[]
\centerline{\includegraphics[width=\columnwidth]{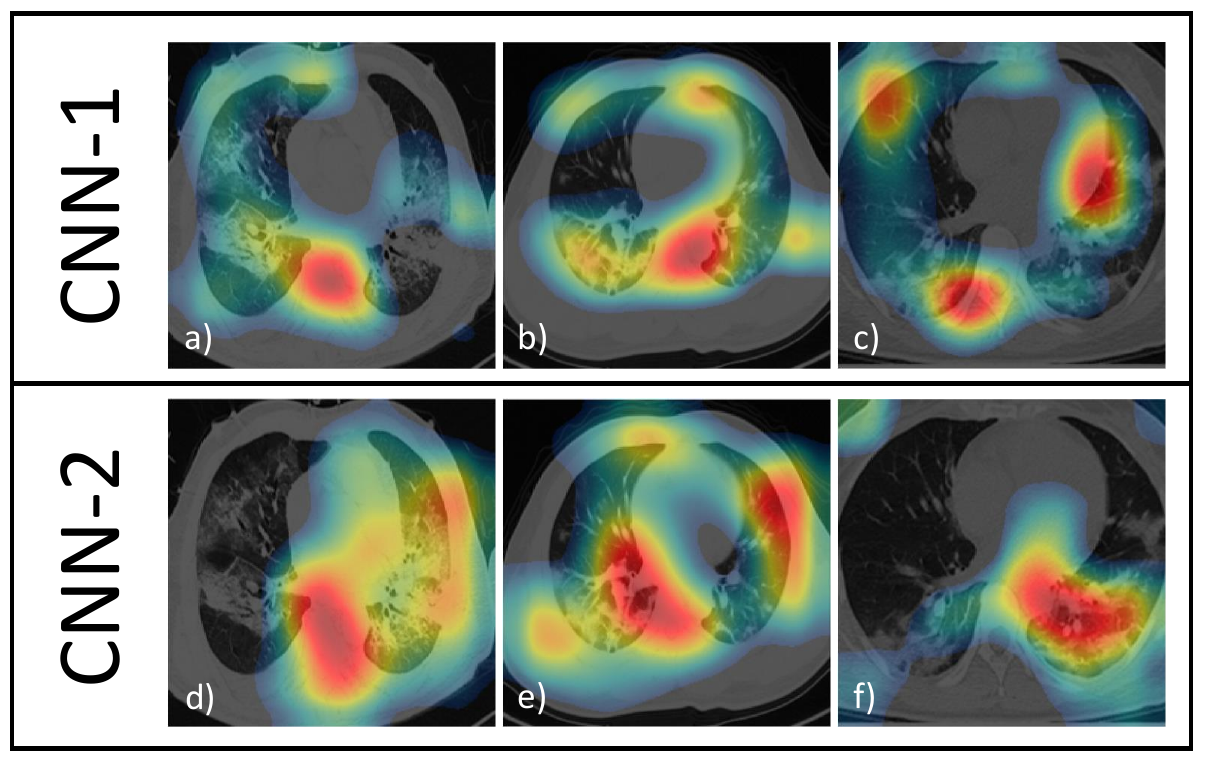}}
\caption{CAMs of CNN-1 and CNN-2 on 3 COVID-19 CT images. Strongest colors (red) implies greater activations. Colors in CAMs are normalized.}
\label{cam1}
\end{figure}

\begin{figure}[]
\centerline{\includegraphics[width=1\columnwidth]{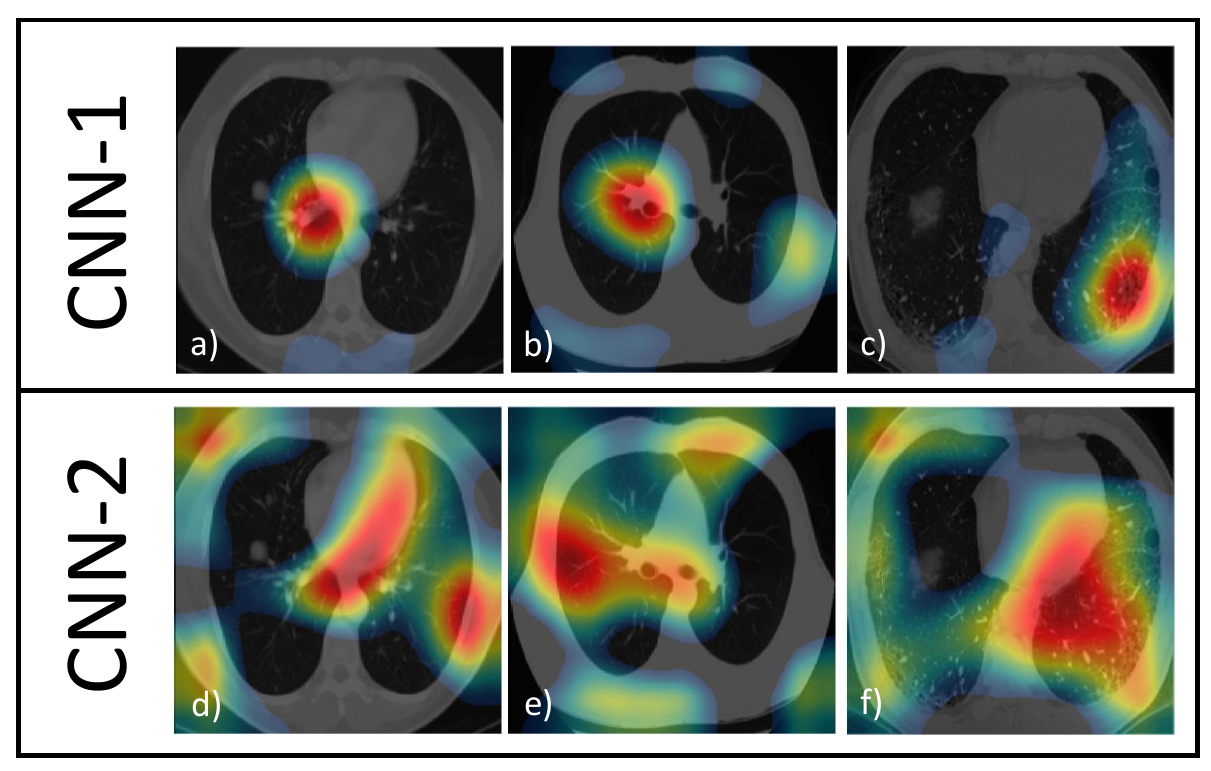}}
\caption{CAMs of CNN-1 and CNN-2 on 3 not COVID-19 CT images. Strongest colors (red) implies greater activations. Colors in CAMs are normalized.}
\label{cam2}
\end{figure}

Figure \ref{cam1} shows 3 examples of CAMs for each CNNs and, to allow comparisons, we refer them to the same 3 CT images (COVID-19 diagnosed both from radiologists and CNNs) extracted from the training dataset. For CNN-1, Figure \ref{cam1}.a, \ref{cam1}.b and \ref{cam1}.c, the activations are not localized inside the lungs. In figure \ref{cam1}.b the activations are just a little bit better than Figures \ref{cam1}.a \ref{cam1}.c, because the red area is partially focused on the ill part of the right lung. The situations enhances in the CAMs of CNN-2 (Figures \ref{cam1}.d, \ref{cam1}.e, \ref{cam1}.f) because the activations are more localized on the ill parts of the lungs (this situation is perfectly represented in Figure \ref{cam1}.f).
Figure \ref{cam2} shows 3 examples of CAMs for each CNNs (as Figure \ref{cam1}) but with 3 CT images of lungs not affected by COVID-19 and correctly classified by both CNNs. CNN-1 focuses on small isolated zones (Figures \ref{cam2}.a, \ref{cam2}.b and \ref{cam2}.c): even if these zones are inside the lungs, it seems unreasonable to obtain a correct classification with so few information (and without having checked the remaining of the lungs). Instead, in CNN-2, the activations seems to take into consideration the whole region occupied by lungs, as demonstrated in Figures \ref{cam2}.d,\ref{cam2}.e and \ref{cam2}.f,  which is the necessary step to correctly classify a lung CT image.

As a conclusion, it is evident that CNN-2 has a better behaviour with respect to CNN-1. Since CNN-1 and CNN-2 have the same model design but different training datatasets, we argue that the training dataset is the responsible of their different behaviour. In fact, the dataset arrangement-2 contains more training images (taken from the italian dataset) and the CNN-2 seems to be gain by it. So, Figure \ref{cam1} and Figure \ref{cam2} suggest that the CNN model, even with a limited number of parameters, is capable to learn the discriminant features of this kind of images. Therefore, the increment of the training dataset should increase also the performance of the CNN.

\begin{table}[]
\center

\caption{Efficiency (calculated as the ratio between sensitivity and number of parameters) comparison between the proposed method and previous works obtained as the ratio between the obtained sensitivity and the number of parameters used to reach it.}
\begin{tabular}{|p{3.5cm}|p{2cm}|p{2cm}|p{2cm}|}
\hline
\textbf{Works} &
  \textbf{\begin{tabular}[c]{@{}c@{}}Sensitivity\\ (\%)\end{tabular}} &
  \textbf{\begin{tabular}[c]{@{}c@{}}\#Parameters\\ (Millions)\end{tabular}} &
  \textbf{\begin{tabular}[c]{@{}c@{}}Sens.  $\div $\\ \#Param. \end{tabular}} \\ \hline
Wang et al. \cite{wang2020deep} & 67    & 23.9 & 2,8   \\ \hline
Xu et all. \cite{xu2020deep}   & 86.7  & 11.7 & 7.41  \\ \hline
Li et al. \cite{li2020artificial} & 90    & 25.6 & 3.52  \\ \hline
The proposed CNN                                       & 85 & 1.26 & 67,46 \\ \hline
\end{tabular}
\label{tab:my-table7}
\end{table}

\subsection{Comparison with recent works}

We compare the results of our work (in particular the CNN-2) with \cite{wang2020deep, xu2020deep, li2020artificial}. Since methods and datasets (training and test) differ and a correct quantitative comparison is arduous, we can have an idea regarding the respective results, summarized in Table \ref{tab:my-table6}. 
The methods \cite{li2020artificial} achieve better results than the method we propose. With respect to \cite{wang2020deep}, our method achieves better results, especially regarding sensitivity which, in our method, is 28\% higher: this suggests a better classification regarding COVID-19 images. 

The average time required by our CNN to classify a single CT image is 1.25 seconds on our high-end workstation. As comparison, the method in \cite{li2020artificial} requires 4.51 seconds on a similar high-end workstation (Intel Xeon Processor E5-1620, GPU RAM 16GB, GPU Nvidia Quadro M4000 8GB). On our medium-end laptot the CNN requires an average time of 7.81 seconds to classify a single image. 
This represents, for the method proposed therein, the possibility to be used massively on medium-end computers: a dataset of about 4300 images, roughly corresponding to 3300 patients \cite{li2020artificial}, could be classified in about 9.32 hours. The improvement in efficiency of the proposed method with respect to the previously compared is demonstrated in Table \ref{tab:my-table7}, where the sensitivity value (the only parameter reported by all the compared methods) is rated with respect the number of parameters used to reach it: the resulting ratio confirms that the proposed method greatly overcomes the others in efficiency.

\section{Conclusion}

In this study, we proposed a CNN design (starting from the model of the SqueezeNet CNN) to discriminate between COVID-19 and other CT images (composed both by community-acquired pneumonia and healthy images). On both dataset arrangements, the proposed CNN outperforms the original SqueezeNet. In particular, on the test dataset the proposed CNN (CNN-2) achieved 83.00\% of accuracy, 85.00\% of sensitivity, 81.00\% of specificity, 81.73\% of precision and 0.8333 of F1Score.  Moreover, the proposed CNN is more efficient with respect to other, more complex CNNs design. In fact, the average classification time is low both on a high-end computer (1.25 seconds for a single CT image) and on a medium-end laptot (7.81 seconds for a single CT image). This demonstrates that the proposed CNN is capable to analyze thousands of images per day even with limited hardware resources. The next major improvements that we want to achieve is to improve the accuracy, sensitivity, specificity, precision and F1Score. In order to do that, since the CNN model seems to be robust as shown with CAMs tests, we aim at increasing the training dataset as soon as new CT images will be available. Moreover, when we compared our methods with those presented in \cite{xu2020deep, li2020artificial} and in \cite{wang2020deep}, we noticed that the last method, as ours, does not use pre-processing, differently from the first two. A possible explanation of the better results of methods \cite{xu2020deep, li2020artificial} with respect to our method could be in the usage of pre-processing. 

As a future work, we aim to study efficient pre-processing strategies that could improve accuracy while reducing computational overhead in order to preserve the efficiency.



\renewcommand{\refname}{\spacedlowsmallcaps{References}} 

\bibliographystyle{unsrt}

\bibliography{refs}


\end{document}

%% file: structure.tex
\usepackage[
nochapters, 
beramono, 
eulermath,
pdfspacing, 
dottedtoc 
]{classicthesis} 

\usepackage{arsclassica} 

\usepackage[T1]{fontenc} 

\usepackage[utf8]{inputenc} 

\usepackage{graphicx} 
\graphicspath{{Figures/}} 

\usepackage{enumitem} 

\usepackage{lipsum} 

\usepackage{subfig} 

\usepackage{amsmath,amssymb,amsthm} 

\usepackage{varioref} 

\theoremstyle{definition} 

\theoremstyle{plain} 

\theoremstyle{remark} 

\hypersetup{
colorlinks=true, breaklinks=true, bookmarks=true,bookmarksnumbered,
urlcolor=webbrown, linkcolor=RoyalBlue, citecolor=webgreen, 
pdftitle={}, 
pdfauthor={\textcopyright}, 
pdfsubject={}, 
pdfkeywords={}, 
pdfcreator={pdfLaTeX}, 
pdfproducer={LaTeX with hyperref and ClassicThesis} 
}